\documentclass[onecolumn,superscriptaddress,amsmath,amssymb]{revtex4-2}

\usepackage{bm}
\usepackage[T1]{fontenc}
\usepackage[english]{babel}
\usepackage[utf8]{inputenc}
\usepackage{longtable}
\usepackage{booktabs} 
\usepackage{comment}
\usepackage{amsmath}
\usepackage{makecell} 
\usepackage[version=4]{mhchem}
\usepackage[colorlinks=true, allcolors=blue]{hyperref}
\usepackage{xspace,graphicx}
\usepackage{multirow}

\usepackage{xcolor}
\def\thefootnote{*}\footnotetext{These authors contributed equally to this work.}

\begin{abstract}
We map the high-pressure phase diagram of nitrogen hydrate up to 16 GPa at room temperature by combining neutron diffraction, Raman spectroscopy, and crystal structure prediction. We reveal a rich sequence of structural transformations, from sI/sII clathrates to hexagonal (sH) and tetragonal (sT) phases, culminating in a previously unknown orthorhombic filled-ice structure above 1.8 GPa in the \textit{Pnma} space group, which we designate as NH-V. This new phase cannot be indexed to any known ice frameworks — such as the high-pressure methane hydrates MH-III (\textit{Imma}) or MH-IV (\textit{Pmcn}) — and exhibits a density approximately 30\% lower than that of stable ice VII, pointing to distinctive water–nitrogen interactions. Our results refine the understanding of nitrogen hydrate behavior under extreme conditions and demonstrate the propensity of \ce{N2} and \ce{H2O} to form stable filled-ice structures up to 16 GPa, with important implications for planetary science.
\end{abstract}

\begin{document}

\title{Discovery of a low-density filled-ice phase in nitrogen hydrate at high pressure}



\author{Selene Berni}
\affiliation{LENS, European Laboratory for Non-Linear Spectroscopy, via Nello Carrara 1, 50019, Sesto Fiorentino (Firenze), Italy}
\affiliation{Dipartimento di Chimica "Ugo Schiff", Universit\`a degli Studi di Firenze, via della Lastruccia 3, 50019, Sesto Fiorentino (Firenze), Italy}
\affiliation{Laboratory of Quantum Magnetism, Institute of Physics,\'{E}cole Polytechnique F\'{e}d\'{e}rale de Lausanne, CH-1015 Lausanne,
Switzerland}

\author{Sophie Espert\thefootnote}
\affiliation{Dipartimento di Fisica, Sapienza Universit\'{a} di Roma, 4 Piazzale Aldo Moro, 00185 Roma, Italy}

\author{Tomasz Poręba\thefootnote}
\affiliation{Laboratory of Quantum Magnetism, Institute of Physics,\'{E}cole Polytechnique F\'{e}d\'{e}rale de Lausanne, CH-1015 Lausanne,
Switzerland}

\author{Simone Di Cataldo}
\affiliation{Dipartimento di Fisica, Sapienza Universit\'{a} di Roma, 4 Piazzale Aldo Moro, 00185 Roma, Italy}

\author{Richard Gaal}
\affiliation{Laboratory of Quantum Magnetism, Institute of Physics,\'{E}cole Polytechnique F\'{e}d\'{e}rale de Lausanne, CH-1015 Lausanne,
Switzerland}

\author{Gabriel Tobie}
\affiliation{Univ Angers, Le Mans Université, CNRS, Laboratoire de Planétologie et Géosciences, UMR 6112, Nantes Université, 2 rue de la Houssinière, Nantes, 44322, France}

\author{Erwan Le Menn}
\affiliation{Univ Angers, Le Mans Université, CNRS, Laboratoire de Planétologie et Géosciences, UMR 6112, Nantes Université, 2 rue de la Houssinière, Nantes, 44322, France}

\author{Thomas C. Hansen}
\affiliation{Institut Laue Langevin, 71 Avenue des Martyrs, Grenoble, France}


\author{Roberto Bini}
\affiliation{Dipartimento di Chimica "Ugo Schiff", Universit\`a degli Studi di Firenze, via della Lastruccia 3, 50019, Sesto Fiorentino (Firenze), Italy}
\affiliation{LENS, European Laboratory for Non-Linear Spectroscopy, via Nello Carrara 1, 50019, Sesto Fiorentino (Firenze), Italy}

\author{Livia Eleonora Bove}

\affiliation{Institut de Min\'{e}ralogie, de Physique des Mat\'{e}riaux et de Cosmochimie (IMPMC), Sorbonne Universit\'{e}, CNRS UMR 7590,  MNHN, 4, place Jussieu, Paris, France}

\affiliation{Dipartimento di Fisica, Sapienza Universit\'{a} di Roma, 4 Piazzale Aldo Moro, 00185 Roma, Italy}

\affiliation{Laboratory of Quantum Magnetism, Institute of Physics,\'{E}cole Polytechnique F\'{e}d\'{e}rale de Lausanne, CH-1015 Lausanne,
Switzerland}

\maketitle

\section{Introduction}

Water exhibits an exceptional diversity of crystalline and amorphous phases depending on thermodynamic conditions, including at least 20 distinct ice polymorphs \cite{Rescigno2025, kuhs2012ice} and several metastable amorphous forms \cite{saltzman2023}. In the presence of guest molecules, water forms complex inclusion compounds-gas hydrates-where small guest species are encapsulated within a hydrogen-bonded lattice \cite{Manakov2017clathratebook}. Among these, clathrate hydrates adopt cage-based architectures stabilized by a delicate balance of van der Waals forces, hydrogen bonding, and hydrophobic interactions. These materials are of interest not only for fundamental studies of intermolecular interactions but also for their roles in geophysics, planetary science, and gas storage technologies \cite{Engelzos1993, Ghosh2019PNAS,Veluswamy2018, Choukroun2012}.

Nitrogen (\ce{N2}), Earth’s most abundant atmospheric component, is known to form clathrate hydrates with water under pressure \cite{kuhs1997cage, chazallon2002situ, champagnon1997nitrogen, pauer1995raman, shoji1982air}. These inclusion compounds are relevant to paleoclimate reconstructions from ice cores, to models of volatile inventories on icy moons and dwarf planets, such as Enceladus, Titan, Europa, or  Pluto\cite{waite2009liquid, lunine1987clathrate, bouquet2015possible, kamata2019pluto, courville2023timing}, and to energy research through guest-exchange processes with methane hydrates \cite{kurnosov2001new}. Nitrogen clathrate hydrate is particularly relevant in the case of Titan which has an dense atmosphere dominated by nitrogen, whose origin is still debated \cite{tobie2014origin}.

At low and moderate pressures, nitrogen hydrate crystallizes in both cubic structure I (sI) and structure II (sII), with sII favored at higher pressures and longer equilibration times \cite{kuhs1997cage, petuya2018guest}. Further compression induces a series of structural transformations: around 0.8 GPa, a hexagonal sH phase emerges, followed by a tetragonal sT phase  \cite{loveday2003structural}. This sequence, reflecting increasing guest–host interactions with pressure, mirrors the behavior seen in other gas hydrates such as hydrogen \cite{ranieri2023} and argon \cite{manakov2001argon, manakov2004structural}. Furthermore, a chiral channel structure (s$\chi$), isostructural to the $C_0$ phase of hydrogen hydrate \cite{delRosso2016}, has been obtained by reloading an empted $C_0$ hydrogen hydrate with nitrogen \cite{Efimchenko2011, kuzovnikov2019, Massani2019}.

\begin{figure} [ht]
\centering
\includegraphics[width=0.8\linewidth]{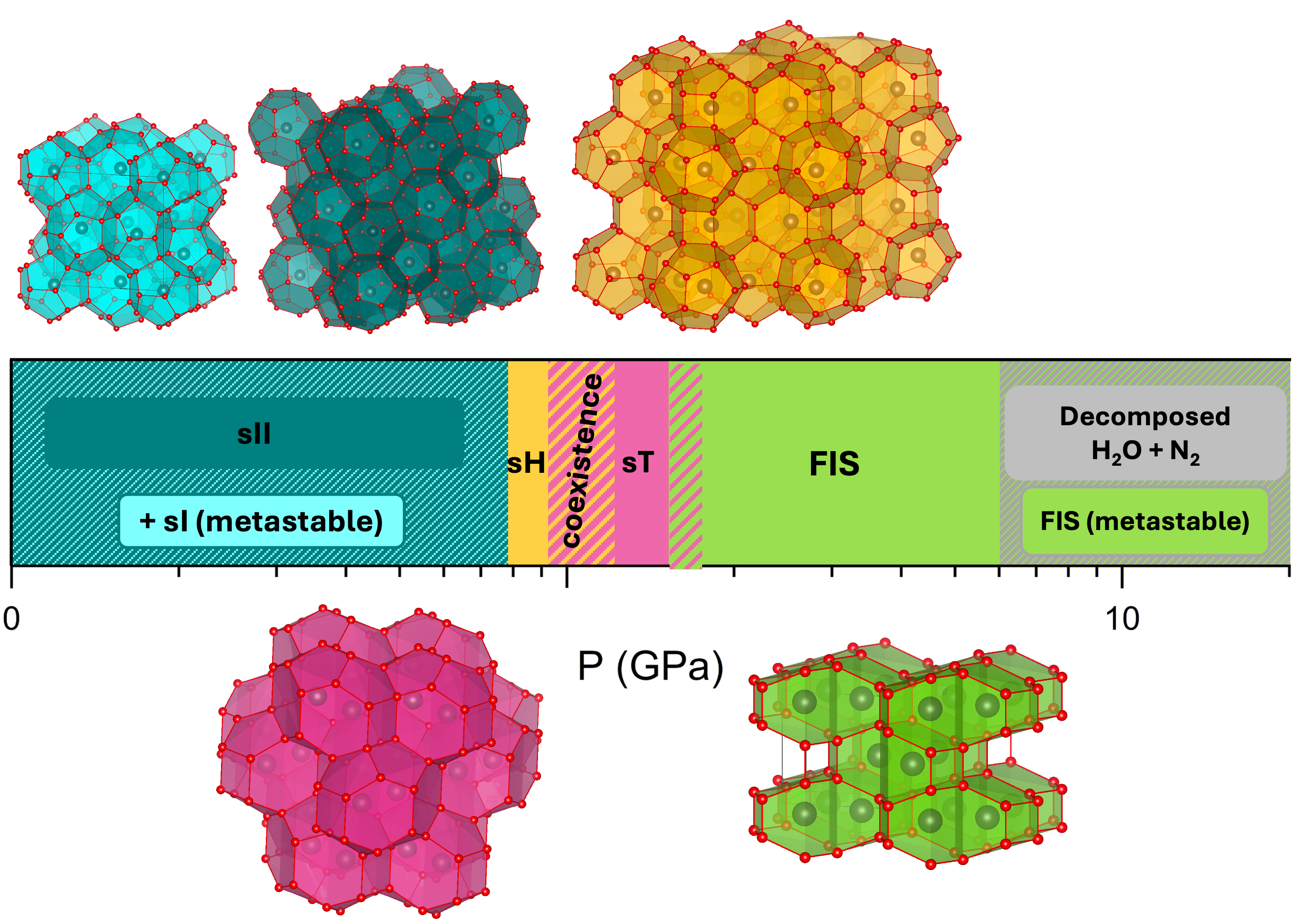}
\caption{\label{fig:diagramma}Structural transition sequence of nitrogen hydrate at ambient temperature as a function of pressure, as observed in this work \textit{via} high-pressure neutron diffraction and Raman spectroscopy.}
\end{figure}

While the stability of nitrogen clathrate hydrates at moderate pressures is well established, the behavior of the water–nitrogen system at higher pressures (> 1.5 GPa), relevant for the deep interiors of large icy moons (Titan, Ganymede and Callisto) and water-rich exoplanets, remains poorly characterized and actively debated. Above 1.8 GPa, the classical clathrate frameworks give way to denser, non-clathrate structures known as filled ices (FIS), in which guest molecules occupy interstitial sites of a hydrogen-bonded water lattice derived from known ice polymorphs \cite{ranieri2023, Poreba2025}. For nitrogen, a filled ice phase, referred to as NH-IV, analogous to methane hydrate III (MH-III) crystallizing in the orthorhombic \textit{Imma} space group, has been proposed \cite{loveday2003structural}. However, published diffraction data show additional Bragg reflections that cannot be satisfactorily explained by this model, suggesting that the high-pressure behavior of the system could be more complex.

The NH-IV phase exhibits a proton-disordered, ice-like framework comprising four-, six-, and eight-membered rings. Although it bears some resemblance to ice Ih in certain crystallographic projections, its topology is sufficiently distorted to accommodate a high nitrogen content, with a \ce{H2O}:\ce{N2} ratio of about 2:1. In comparison, the preceding sT phase contains fewer guest molecules ($\sim$3:1), indicating significant nitrogen enrichment during the transition to the filled ice state—a behavior also observed in high-pressure hydrogen \cite{Poreba2025} and methane \cite{Schaack2018} hydrates. Raman spectroscopy by Sasaki et al. further suggested the appearance of an additional distinct filled ice phase between 2 and 6 GPa \cite{sasaki2003microscopic}.

Whether stable \ce{N2}–\ce{H2O} compounds persist to even higher pressures remains controversial. Zhang et al. \cite{zhang2021} proposed that above 6 GPa, the nitrogen–water system becomes immiscible, in contrast to the remarkable stability of hydrogen and methane hydrates up to megabar pressures \cite{Schaack2019, ranieri2023}. In this regime, nitrogen was observed as fine-grained aggregates embedded in the ice matrix, rather than as a crystalline compound. To date, no Rietveld refinements have been reported above 1.5 GPa, leaving open the question of whether well-defined molecular \ce{N2}–\ce{H2O} structures exist beyond the clathrate regime.

In this work, we present a comprehensive study of the high-pressure phase diagram of nitrogen hydrate, combining neutron diffraction, Raman spectroscopy, and structural prediction to probe pressures up to 16 GPa at room temperature and down to 80 K at 2.1 GPa. Our results uncover a rich sequence of structural transitions, clarifying the stability and coexistence domains of known phases and revealing a previously unreported filled ice phase above 1.8 GPa, which cannot be indexed within the NH-IV framework. We designate this new phase NH-V and show that it remains stable up to 5.2 GPa at both room and low temperatures, before transforming into the known NH-IV phase at higher pressures.

The pressure–temperature stability field of NH-V is particularly relevant to the interiors of icy moons such as Titan, Ganymede, and Callisto, with implications for their volatile budgets, redox conditions, and potential habitability. Furthermore, structural analogies among O\textsubscript{2}-, H\textsubscript{2}-, and N\textsubscript{2}-filled ices suggest possible surface co-localization and enhanced reactivity of these species, linking hydrate physics to planetary geochemistry.

Figure~\ref{fig:diagramma} summarizes the current understanding of the nitrogen hydrate phase evolution up to 16 GPa at room temperature, combining previous studies \cite{sasaki2003microscopic, loveday2003structural} with our new data. This diagram provides a reference framework for the discussion that follows, where we detail the structural nature of the newly observed NH-V filled ice phase and the transition region between clathrate and filled ice regimes.

\section{Results}

In order to follow the pressure evolution of nitrogen hydrate, hydrogenated \ce{N2}–\ce{H2O} samples (prepared as described in the Methods section) were cryoloaded into a diamond anvil cell and cycled between 0.4 and 16 GPa while collecting in situ Raman spectra.

\begin{figure*} [ht]
\centering
\includegraphics[width=0.8\columnwidth]{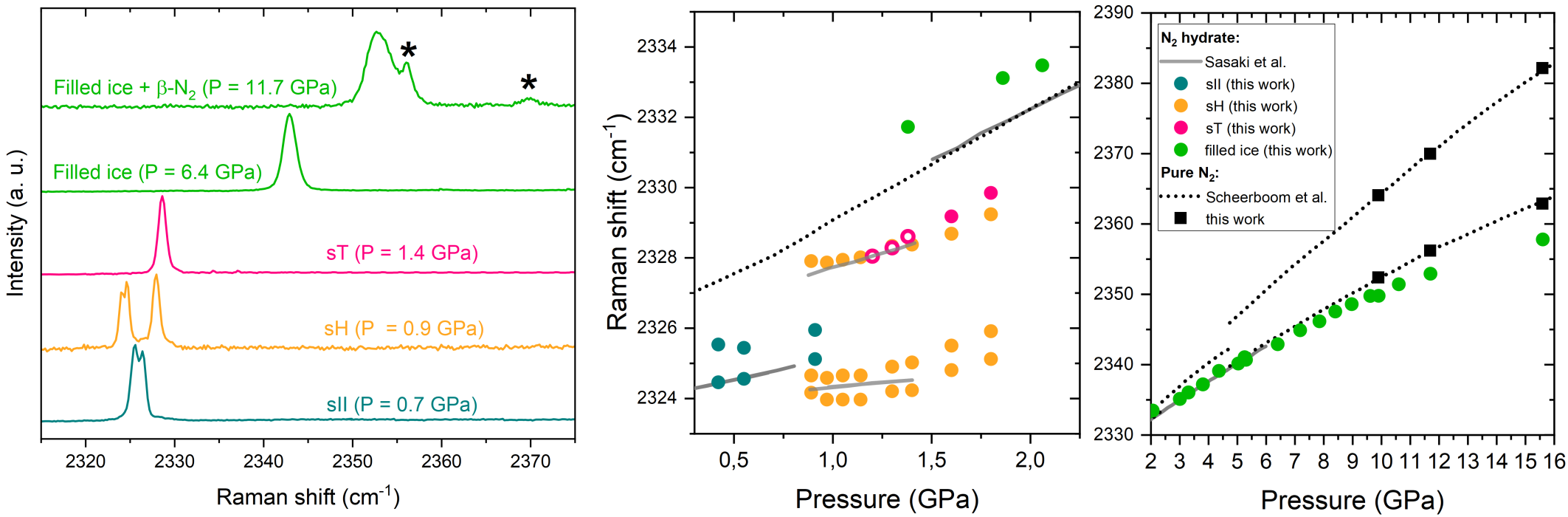}
\caption{\label{fig:raman} Left: Raman spectra of the N$\equiv$N stretching bands of nitrogen hydrate. The pressure at which the spectra were acquired is reported beside each spectrum. In the top spectrum, the pure \ce{N2} present in some parts of the samples together with \ce{N2} filled ice is marked with two asterisks.
Center and right: Raman shifts of pure and hydrated N$\equiv$N stretching bands. The data represented by circles are from this work. As indicated in the legend, data from the sII phase are reported in blue, sH phase in yellow, sT phase in pink, and the filled ice phase in green. The full circles are from data acquired upon compression, while the empty ones come from acquisitions upon decompression. With black squares we reported the pure \ce{N2} present in some parts of our samples together with \ce{N2} filled ice. 
The pressure evolution of pure nitrogen stretching bands are from Sheerboom et al. \cite{scheerboom1996high} and are reported with small dots, while the full gray lines are relative to the trend of the nitrogen hydrate studied by Sasaki et al. \cite{sasaki2003microscopic}.}
\end{figure*}

Figure~\ref{fig:raman} shows representative Raman spectra of the \ce{N#N} stretching band across the various hydrate phases (left), the pressure evolution of the Raman shift (center, right), and comparison with literature data for pure nitrogen \cite{scheerboom1996high} and previously measured nitrogen hydrates \cite{sasaki2003microscopic}. Spectral features were assigned based on phase-specific trends and cross-validated with diffraction data.

\begin{figure*}
\centering
\includegraphics[width=0.8\columnwidth]{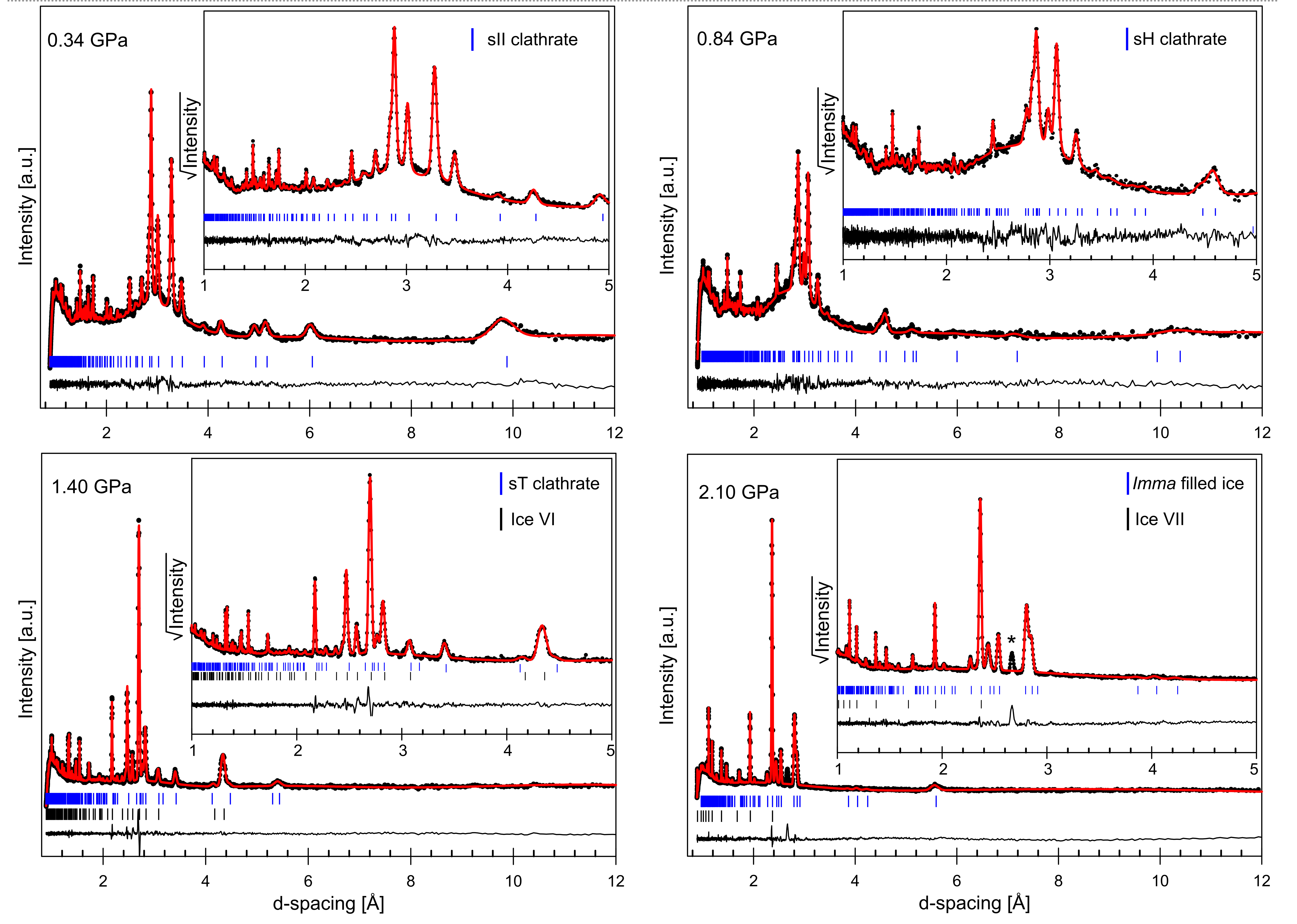}
\caption{\label{fig:diffraction}Neutron diffraction patterns of a powder sample of deuterated nitrogen hydrate collected upon compression in a PE press on D20 beamline of ILL ($\lambda$ = 1.54 \AA). The red line is the  refined pattern using Le Bail fit with Topas software, while the data are reported in black.}
\end{figure*}

At ambient temperature and 0.3 GPa, diffraction data (Figure \ref{fig:diffraction}, top left) confirm the presence of the cubic sII clathrate (Fd$\bar{3}$m), with refined lattice parameter $a = 17.117(1)$ \AA. Raman spectra in this range display two \ce{N#N} stretching bands at 2324.5 and 2325.5 cm$^{-1}$, attributed to nitrogen molecules in large and small cages, respectively. Below 1 GPa, spectra in some parts of the samples showed only a single stretching band—consistent with Sasaki et al. \cite{sasaki2003microscopic}—likely reflecting incomplete crystallization or local disorder during nucleation. Neutron diffraction (Figure SI-1) also detected metastable sI-type clathrate ($a = 11.824(2)$ \AA) after loading at 0.6 GPa, suggesting sI-to-sII conversion may not have been fully complete. The coexistence of sI (Pm$\bar{3}$n) and sII (Fd$\bar{3}$m) and nonuniform cage filling may account for anomalies in the low-pressure Raman spectra, including deviations from the expected large-to-small cage intensity ratios.

Upon compression to 0.8 GPa, diffraction and Raman data indicate a transition to the hexagonal sH phase (\textit{P}6/\textit{mmm}), with refined lattice parameters $a = 11.993(2)$ \AA~and $c = 9.936(5)$ \AA~ (Figure \ref{fig:diffraction}, top right). Raman spectra at this pressure exhibit three bands ($\sim$2328–2329.5 cm$^{-1}$), assigned to nitrogen in distinct cage sites (L, S, and S' cages). Broad diffraction features indicate residual excess water.

At 0.9 GPa, a tetragonal sT phase (\textit{P}4$_2$/\textit{mnm}) appears, coexisting with sH and ice VI (Figure SI-2). The sT phase, better identified upon decompression in the Raman spectra due to the absence of coexisting sH phase, exhibits a sharp single \ce{N#N} Raman band around 2329 cm$^{-1}$, consistent with its single-cage topology (see pink dots in Figure \ref{fig:raman}). Le Bail refinement at 1.4 GPa yields $a = 6.312(1)$ \AA~, $c = 10.578(10)$ \AA~.

\begin{figure}
\centering
\includegraphics[width=1 \linewidth]{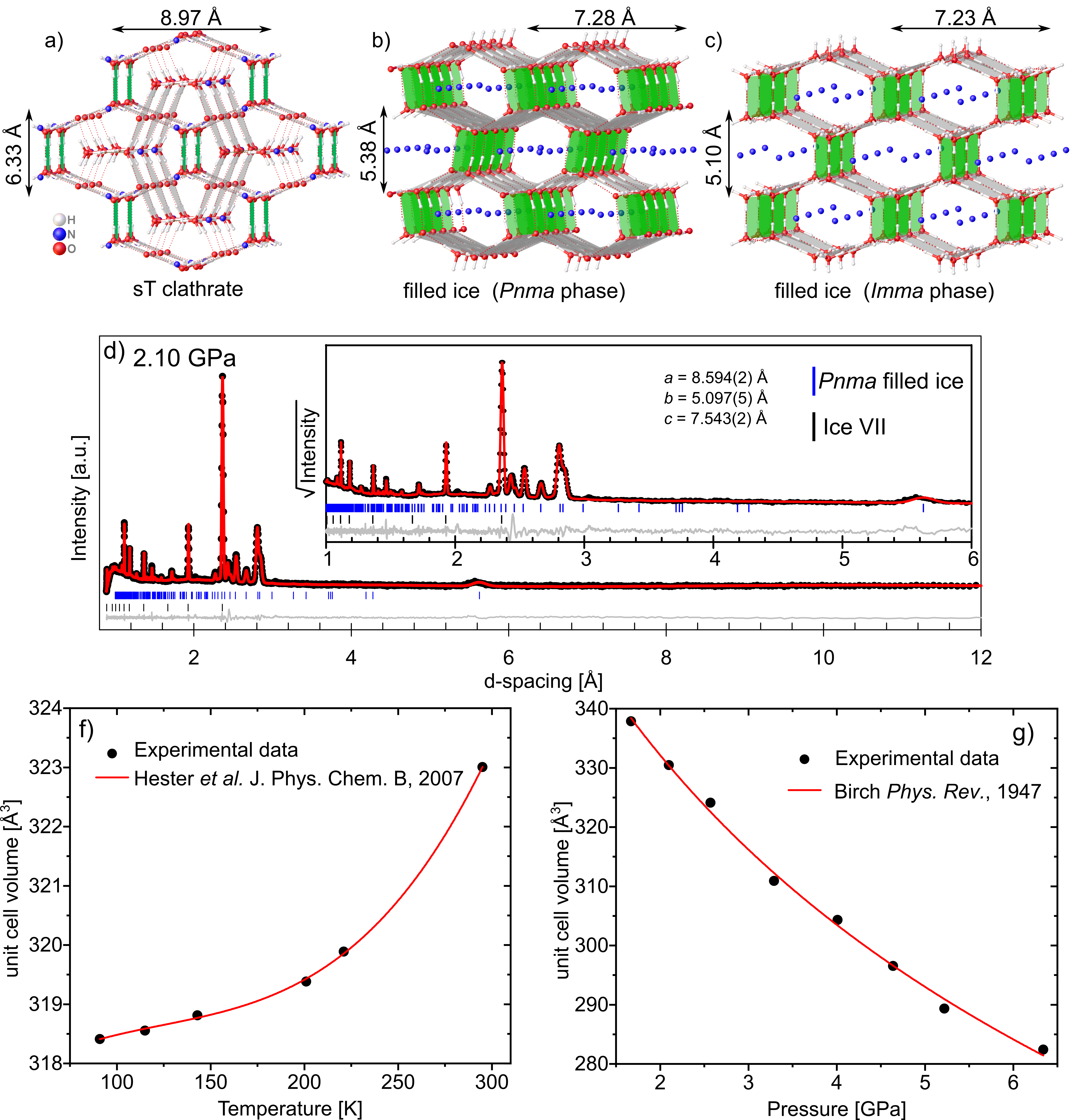}
\caption{\label{fig:sequenza} 
Top panel: structural evolution sequence from (a) sT (NH III) to (b) \textit{Pnma} (NH-V) to (c) \textit{Imma} (NHV). 
Medium panel (d): Example of Le Bail refinement of the 2.1 GPa diffractogram of the new \textit{Pnma} structure. The lattice parameters are indicated in the legend. Some ice VII was also present in the sample (see text).
Bottom panel: (f) Temperature dependence of the unit cell volume of NH-V at 2.6 GPa, measured during isobaric heating from 91 to 300 K. Data were fitted with the thermal expansion model of Hester \textit{et al.} \cite{hester2007thermal}. (g) Pressure dependence of the unit cell volume of NH-V, obtained from Le Bail refinements of neutron diffraction patterns between 2 and 7 GPa. The red line represents the fit using a third-order Birch–Murnaghan \cite{birch1947finite} equation of state. }
\end{figure}

Above 1.8 GPa, the clathrate framework collapses and diffraction patterns collected at 2.1 GPa (Figure \ref{fig:diffraction}, bottom right) show a transition to a new phase. Initial fits using the orthorhombic \textit{Imma} model (NH-IV) reported by Loveday et al. did not fully account for an extra Bragg peak at $d = \sim 2.7$ Å, also observed in prior studies \cite{loveday2003structural}. Previous works on methane hydrate (largely isostructural to nitrogen hydrate at high pressure) also reported the occasional presence of a metastable primitive orthorhombic phase upon compression to 1.8 GPa \cite{Hirai2002}. This suggests a distinct polymorph, here designated as NH-V. Refinement using a primitive orthorhombic cell (\textit{Pnma}) gives $a = 8.594(2)$ \AA, $b = 5.097(5)$ \AA, $c = 7.543(2)$ \AA. Evolutionary structure search calculations support a \textit{Pnma} model, within 1 meV/atom from the \textit{Imma} phase at 2.1 GPa, with a 2:1 \ce{H2O}:\ce{N2} ratio and hydrogen bond motifs similar to high-pressure methane hydrates \cite{Hirai2002}. Structural analysis reveals a collapse of the 4\textsuperscript{2}5\textsuperscript{8}6\textsuperscript{4} sT cages into octagonal, nitrogen-filled 1D channels constructed from cross-linked water hexamers and windowpane-like tetramers, highlighting the topological relationship between clathrate and filled-ice structures (Figure \ref{fig:sequenza}). 
The NH-V (\textit{Pnma}) phase remains stable up to $\sim$5.2 GPa before transforming into the more symmetric NH-IV (\textit{Imma}) phase, which persists under further compression. The  NH-IV phase can also be recovered metastably upon decompression to 2 GPa (Figure SI-3).
Table \ref{tab:your_label} summarizes the crystallographic and compositional data for all identified phases.
\renewcommand{\arraystretch}{1.5}
\begin{table}
  \centering
    \caption{ Unit cell parameters and stoichiometry of the different phases of nitrogen hydrate identified at high pressure.}
  \begin{tabular}{|c|c|c|c|c|c|}
    \hline
    & sII & sH & sT & \textit{Pnma} FIS & \textit{Imma} FIS \\
    \hline
    Pressure [GPa] & 0.3 & 0.9 & 1.4 & 2.1 & 5.2 \\
    \hline
    Space group & $Fd\Bar{3}m$ & $P_6/mmm$ & $P4_2/mnm$ & $Pnma$ & $Imma$ \\
    \hline
    Lattice parameters [\AA] &
    \makecell{\textit{a}=17.117(1)} &
    \makecell{\textit{a}=11.993(2) \\ \textit{c}=9.936(3)} &
    \makecell{\textit{a}=6.312(1) \\ \textit{c}=10.578(10)} &
    \makecell{\textit{a}=8.594(2) \\ \textit{b}=5.097(1) \\ \textit{c}=7.543(2)} &
    \makecell{\textit{a}=7.840(2) \\ \textit{b}=4.852(2) \\ \textit{c}=7.591(2)} \\
    \hline
    Unit cell volume [\AA$^{3}$] & 5015.1(8) & 1237.9(6) & 421.5(8) & 330.5(1) & 289.3(3) \\
    \hline
    Unit cell composition & 24 \ce{N2} $\cdot$ 136 \ce{H2O} & 7 \ce{N2} $\cdot$ 34 \ce{H2O} & 4 \ce{N2} $\cdot$ 12 \ce{H2O} & 4 \ce{N2} $\cdot$ 8 \ce{H2O} & 4 \ce{N2} $\cdot$ 8 \ce{H2O} \\
    \hline
    \ce{H2O} to \ce{N2} ratio & $\approx5.66:1$ & $\approx4.85:1$ & 3:1 & 2:1 & 2:1 \\
    \hline
  \end{tabular}
  \label{tab:your_label}
\end{table}

Raman spectra confirm that NH-IV (\textit{Imma}) remains stable up to 16 GPa under slow compression (0.1 GPa/h), exceeding previous stability estimates \cite{sasaki2003microscopic} and suggesting greater structural adaptability compared to \ce{O2} \cite{frost2025implications} and \ce{Ar} hydrates \cite{marounina2015evolution}. After two weeks at 10 GPa and room temperature, however, NH-IV decomposed into pure \ce{N2} (evident from the $\delta$-phase doublet in Raman \cite{bini2000high, scheerboom1996high}), and ice VII, indicating its metastability over long timescales—similar to \ce{H2} hydrates \cite{ranieri2023}.

To quantify the compressibility of the NH-V phase, we fitted the pressure–volume relationship derived from neutron diffraction data between 2 and 7 GPa using a third-order Birch–Murnaghan isothermal equation of state (Figure~\ref{fig:sequenza} bottom panel a) \cite{birch1947finite}, assuming the orthorhombic \textit{Pnma} symmetry. The resulting fit yielded a bulk modulus $B_0= (10.3 \pm 3.4)$  GPa, a pressure derivative $B'= 4.1 \pm 1.0 $ , and a reference volume $V_0= (383.5 \pm 10.7)$  \AA$^3$ (R$^{2}$=0.99686) (for comparison the ice VII bulk modulus is 20.77 GPa \cite{bezacier2014equations}).

A second hydrate sample was compressed above 2 GPa, showing again sT direct transition into the lower symmetry (\textit{Pnma}) NH-V instead of (\textit{Imma}) NH-IV, confirming the sequence of transition. To examine the thermal stability of NH-V, the sample was cooled to 90 K at 2.6 GPa and then heated isobarically to 300 K, collecting diffraction patterns throughout (Figure~\ref{fig:sequenza}g). The Le Bail refinement of the 91 K data yielded cell parameters of $a = 8.548(39)$ \AA, $b = 5.039(30)$ \AA, and $c = 7.382(32)$ \AA.
The thermal expansion behavior was modeled using the semi-empirical expression proposed by Hester \textit{et al.} \cite{hester2007thermal}, replacing the lattice parameter $a_0$ with volume $V_0$ at 91 K (318.4 \AA$^{3}$). The values of the fit, provided in Figure~\ref{fig:sequenza}f, resulted in $V_1= (2.7\pm 0.5)10^{-5}$, $V_2=(-4.1\pm 1.5)10^{-7}$, $V_3=(6.1\pm 0.7)10^{-9}$ (R$^{2}$=0.9996). When comparing the relative expansion of NH-IV with literature data on \ce{N2}, \ce{CH4}, and \ce{CO2} clathrates (see Figure SI-4), it appears that NH-V expands less than typical sI or sII clathrates' structures. Instead, its expansion closely mirrors that of deuterated ice I\textsubscript{h}, consistent with the structural relationship between NH-V and the distorted hexagonal ice network framework \cite{loveday2001transition}.

\begin{figure} [ht]
\centering
\includegraphics[width=1\linewidth]{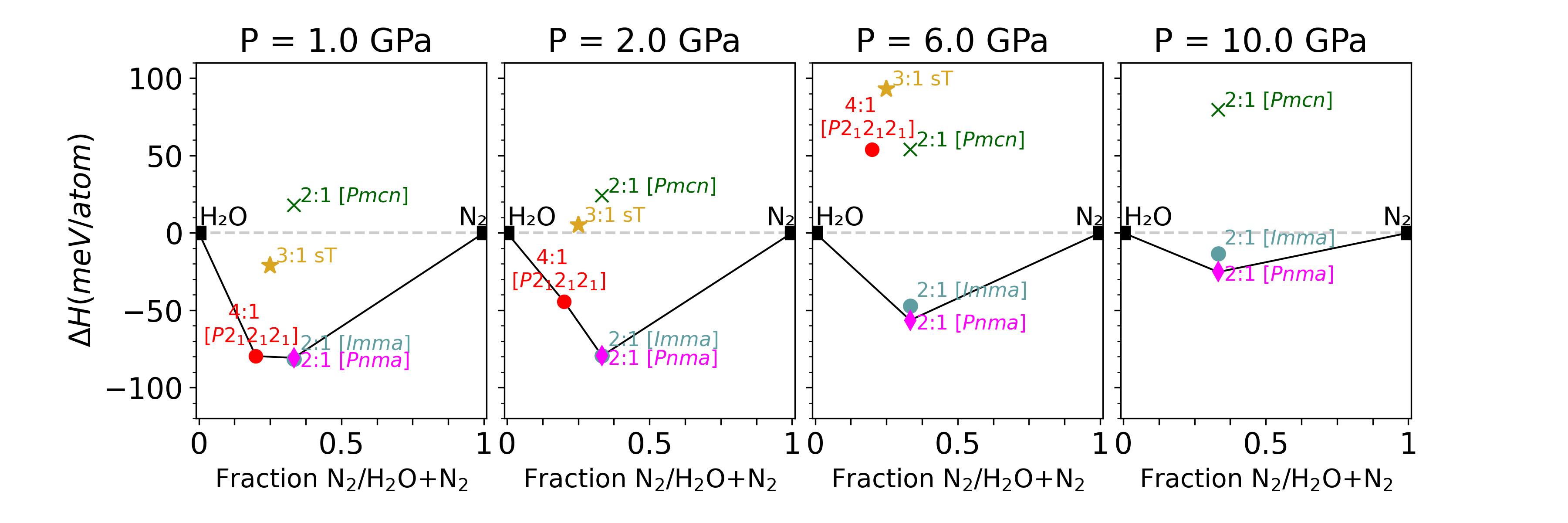}
\caption{\label{fig:convex} Convex hull diagrams of the \ce{H2O}–\ce{N2} system at 1, 2, 6 and 10 GPa, showing the formation enthalpy relative to ice and solid \ce{N2} of the most relevant crystal structures, labeled with their space group. Thermodynamically stable phases lie on the convex hull.}
\end{figure}

To further investigate the nature and thermodynamic stability of the newly observed high-pressure phase NH‑V, we performed a combination of fixed- and variable-composition structure prediction calculations using the evolutionary algorithm as implemented in the USPEX code, constrained to \ce{H2O} and \ce{N2} molecules \cite{Zhu2012} and enthalpies were computed using Quantum ESPRESSO. Structure prediction calculations were performed at 2 and 4 GPa for the 2:1, 4:1, 5:1, 6:1 and 5:3 composition. The most stable crystal structures found at each pressure were then relaxed as a function of pressure. The formation enthalpies relative to pure ice and solid nitrogen are reported in the convex hull diagrams at 1, 2, 6, and 10 GPa (Figure~\ref{fig:convex}), where we only show the structures with compositions 4:1 , 3:1 and 2:1, relevant to the discussion. 

The convex hull diagram (black line in Figure~\ref{fig:convex}) indicates which structures are thermodynamically unstable with respect to decomposition into other phases, i.e. points lying on the hull. Structures above the hull would decrease their internal energy by decomposing into stable structures, although the diagram does not take into account the system's kinetics, hence the potential for metastability.

At 1 GPa, a structure with \textit{P2\textsubscript{1}2\textsubscript{1}2\textsubscript{1}} space group and 4:1 \ce{H2O}:\ce{N2} (denoted NH 4:1) composition was found to lie on the convex hull. However, NH 4:1 becomes unstable above $\sim$3 GPa, suggesting it has a narrow stability range. This structure was never observed in our diffraction patterns even as metastable.

At 2 GPa, we identified two nearly-isoenergetic phases (within 5 meV/atom) with 2:1 \ce{H2O}:\ce{N2} composition. One is the \textit{Imma} phase proposed by Loveday et al. \cite{loveday2003structural}, and the other is a previously unreported phase with \textit{Pnma} space group. This second phase is consistent with our diffraction patterns.

The two phases remain nearly isoenergetic between 2 and 5 GPa, and both become progressively less stable at higher pressures. The two structures, NH‑IV (\textit{Imma}) and the newly assigned NH‑V (\textit{Pnma}), are likely both accessible in the intermediate 1.8 - 6 GPa range, with the \textit{Pnma} phase probably favored at room temperature and under non‑fully hydrostatic compression due to its more distorted framework. 

We also manually included two other crystal structures to the convex hull. The first is a \textit{Pmcn} phase with 2:1 \ce{H2O}:\ce{N2} composition that is isostructural to MH-IV \cite{Schaack2019}. This structure exhibits a positive formation enthalpy, and is over 100 meV/atom above the convex hull - strong evidence that it is unlikely to form. The sT structure was obtained from the experimental CIF file of argon hydrate \cite{manakov2001argon} by replacing the argon guest with an \ce{N2}  molecule positioned at the center of the cages. No further optimization of the \ce{N2} orientation or of the surrounding \ce{H2O} framework positions was performed, which may account for the discrepancy between the calculated stability pressure and the experimental observations. Note also that a possible pressure shift can occur between experiments and simulations, where the sT phase appears stable at pressures below 0.1 GPa. The structure is unstable at 1 GPa, as is 60 meV/atom above the hull. This may be attributed also to limitations in DFT accuracy, particularly in the estimation of relative enthalpies under pressure.

At 2 GPa, the compositions on the convex hull remain the same as 1 GPa, although the 4:1 \textit{P2\textsubscript{1}2\textsubscript{1}2\textsubscript{1}} phase is almost unstable, and it becomes unstable at 6 GPa, leaving only the 2:1 \textit{Pnma} phase on the hull. As pressure further increases, so does the stability of the 2:1 \textit{Pnma} phase which, according to our calculations, becomes unstable at about 13 GPa. Experimentally, the qualitative trend is similar, although decomposition appears to occur around 6 GPa rather than the 13 GPa predicted by our calculations. This discrepancy may arise from the known limitations of the PBE functional, which tends to underestimate the energy of the strongly bound N-N triple bond and may over-stabilize nitrogen-rich compounds under pressure.

In summary, the sT → \textit{Pnma} →  \textit{Imma} sequence is maintained, although the transition pressures in the simulations may differ from those observed experimentally. This discrepancy is due in part to the simulations being performed at zero temperature.

\begin{figure}
\centering
\includegraphics[width=0.7\linewidth]{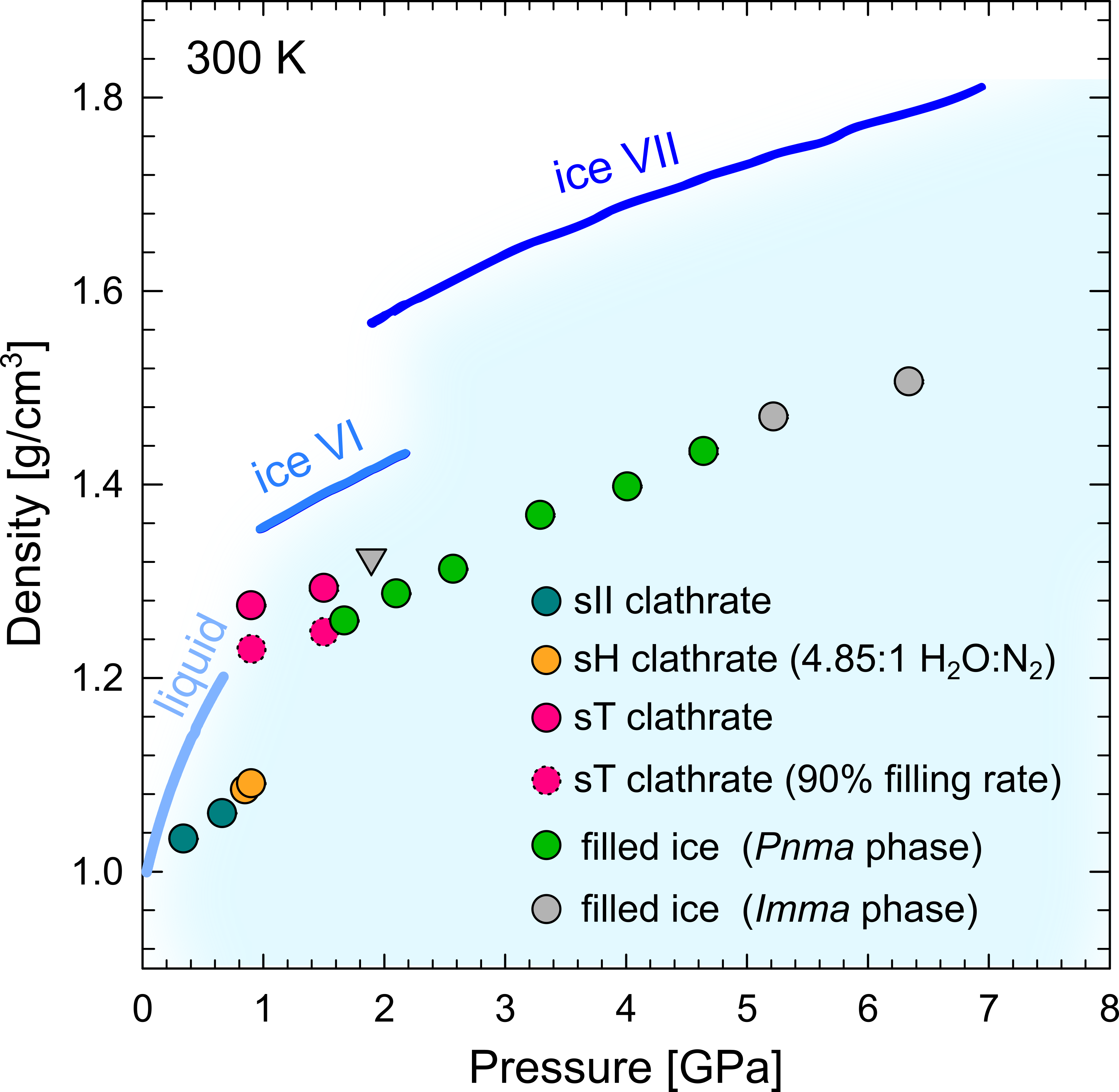}
\caption{\label{fig:densities} Densities of high-pressure phases of nitrogen hydrate based on the unit-cell volumes derived via neutron diffraction, assuming full nitrogen occupancy. Blue lines represent experimentally-derived densities of liquid water, ice VI and ice VII, respectively (Ref \cite{Grindley1971} and \cite{Fortes2012}). Circles and a triangle denote data collected upon compression and subsequent decompression, respectively.}
\end{figure}

The evolution of the density of the nitrogen hydrate phases, shown in Figure \ref{fig:densities}, reflects the progressive densification upon compression, while remaining consistently below the densities of pure ice VI/VII under comparable conditions. The transition from sII to sH is accompanied by a modest density increase, followed by a sharper jump ($\sim$ 18.5\%) at the sT transition, consistent with increased nitrogen uptake. However, precise refinement of nitrogen occupancy in the sT phase was not possible due to its narrow observation range under compression and peak overlap with coexisting ice VI. As a consequence a nominal full cage occupancy was assumed. This likely overestimates the true density of sT: a slightly lower occupancy (90\%) yields densities that better match the expected monotonic densification trend. Notably, the subsequent transition from a fully occupied sT phase to NH‑V phase exhibits a slight density drop (2.8\%), which may reflect incomplete cage or channel filling, as observed in other gas hydrates \cite{manakov2004structural}. Upon decompression, the expected density trend is recovered, highlighting kinetic barriers to achieving full occupancy in the compressed filled-ice channels.

Finally, the low-frequency Raman bands between 2 - 10 GPa (Figures SI-5, SI-6, SI-7) exhibit broad phonon features consistent with partially ordered filled-ice structures, matching signatures reported for filled-ice methane hydrates \cite{scelta2022addressing}.

\section{Conclusions}
In conclusion, we have clarified the high-pressure phase behavior of nitrogen hydrate up to 16 GPa using a combined approach of neutron diffraction, Raman spectroscopy, and structure prediction. Our results trace the full sequence of transformations from open clathrate frameworks (sII, sI, sH, sT) to filled-ice structures NH-V (Pnma) and NH-IV (Imma), revealing how guest–host interactions, structural rearrangements, and density changes are closely interlinked.

A key outcome of this study is the identification and structural characterization of a previously unresolved filled-ice phase in the Pnma space group (NH-V), which forms above 1.8 GPa as the sT clathrate collapses into a more compact, channel-based topology. This intermediate structure, less dense than the Imma phase that follows, highlights the non-monotonic path of densification in the nitrogen–water system—shaped not only by thermodynamics but also by kinetic and configurational constraints.

The stability range of NH-V overlaps with the expected pressure–temperature conditions in the deep interiors of icy moons, suggesting that such metastable nitrogen-rich hydrates could affect internal stratification, volatile storage, and chemical evolution. During the first billion years after accretion,  the progressive warming of the deep interiors of these large icy moons, consisting of a mixture of organics, ices and rock \cite{tobie2014origin}, is expected to result in a progressive release of nitrogen-bearing compounds, first ammonia and then molecular nitrogen as the interior warms up (T > 500 K) \cite{bruch, moore}. Our results show that any produced nitrogen molecule will interact with water molecules as they migrate upward and will be stored temporarily in the form of NH-V, and then progressively transform in the different phases as they decompressed during their transport toward the outer ice shell. The temporary storage of nitrogen in gas hydrates and their subsequent slow transport through the hydrosphere would allow a relatively late formation of Titan's atmosphere. This is consistent with the late formation scenario proposed by \cite{marounina2015evolution, Neri2020}, to explain the presence of a massive atmosphere on Titan despite the strong erosion induced by massive impacts during the Late Heavy Bomdardment \cite{gomes2005origin}. Future modeling work is needed to quantify the timing of nitrogen production by organic alteration, storage and transport to the outer layer and finally out-gassing at the surface. 

By resolving a previously open structural question, this work contributes to a clearer picture of how gas hydrates adapt under pressure and what role they might play in shaping planetary environments.

\section{Methods}

\paragraph{Sample preparation}
The sI/sII phase nitrogen clathrate hydrate was prepared by means of \textit{ex situ} synthesis following a well assessed procedure reported for various clathrates \cite{chazallon2002situ, scelta2022addressing, berni2023high, berni2024exploring}.
The \ce{D2O} or \ce{H2O} ice powder was produced by nebulization of MilliQ water in a liquid nitrogen bath using a SinapTec 80Khz (NexTgen Inside 80kHz 30 W), known to produce ice grains of 90 µm size.
The condensed ice was then sifted through a sieve.
In the case of deuterated water, the procedure was performed under  nitrogen atmosphere to avoid the condensation of non-deuterated atmospheric water.
2 g of water ice powder were placed in an autoclave kept at -30°C and inflated with 160 bar of the desired gas. The temperature was slowly increased to +0.5°C, where the enclathration is supposed to take place. Several temperature cycles around 0°C were performed.
The autoclave was opened in a cold chamber (-25°C). The clathrate was tested by means of Raman spectroscopy with a Qontor Raman spectrometer at ambient pressure in a liquid nitrogen continuous-flow cryostat (Oxford instruments) with optical access. Samples showing a good amount of enclathrated gas were transferred in vials and stored in dewars under liquid nitrogen.

\paragraph{Loading procedures}
Concerning spectroscopic characterization, screw-driven diamond anvil cells (Mao-Bell design, DAC) with 500 $\mu$m culet diamonds were used to generate pressures of the order of gigapascals.
Preindented stainless steel gaskets with 150 $\mu$m  holes were used to contain the sample.
Ruby balls were placed in the sample chamber to probe the pressure using the ruby photoluminescence method \cite{mao1986calibration}.
The high-pressure cells were loaded while in a bath of liquid nitrogen.
Once the cell was closed, it was removed from the liquid nitrogen bath and pressure was applied by tightening the screws.

\paragraph{Reman spectroscopy}
Raman spectra of the pressurized samples were collected with LabRaman HR (exciting wavelength of 532 nm, nominal power of 1 W), both with a 600 lines/mm grating and a 1800 lines/mm one (only with the 1800 lines/mm grating it was possible to resolve the splitting of the nitrogen band at low pressure) and fitted with Fityk software \cite{wojdyr2010fityk} using Voigt functions.
Lower wavenumbers Raman spectra were obtained with a \textit{in-house} set-up in back-scattering geometry, equipped with a 660 nm line of a diode laser as an excitation source and using a long working distance 20× Mitutoyo micro-objective. Spectra were collected thanks to an Acton/SpectraPro 2500i monochromator equipped with holographic super notch filters and a CCD detector (Princeton Instruments Spec-10:100BR). Such set-up allowed a spatial resolution on the sample of about 3 $\mu m$ and a spectral resolution of 1 cm$^{-1}$.

\paragraph{Neutron diffraction}
Around  30 mm\textsuperscript{3} of the deuterated sample was cryoloaded in a spherical TiZr gasket and sealed with approximately 3 kbar in the clamp modulus, subsequently transferred to a Paris-Edinburgh (P-E) press coupled with a cryostat \cite{Klotz2016}. The press was cooled by two-stage closed-cycle refrigerator in an “orange” cryostat. A lead chip was added to the sample to monitor the pressure following its equation of state \cite{strassle2014equation}. Two loadings were performed: the first sample was slowly compressed up to 6 GPa, while the second one was compressed to about 2 GPa and then cooled down to 100 K and slowly heated up to ambient temperature. The neutron diffraction data were collected at the high-intensity diffractometer D20 at the Institut Laue-Langevin (Grenoble, France) \cite{Hansen_2008}, using a wavelength of 1.54 \AA{} \cite{Berni2024ILL}.
The diffraction patterns used for the compressibility study were acquired for two hours after compression. Due to limited acquisition time, measurements for the thermal expansivity study were made during continuous heating; thus, the results reflect non-equilibrated conditions. The sequential refinements tracked unit cell evolution, and all volumes were normalized to 2.6 GPa using the isothermal equation of state obtained from the compressibility study. The temperature was measured at the top of the anvil cell throughout the acquisition of the diffraction pattern.
Unit cell parameters were obtained by Le Bail fitting of neutron diffraction patterns using TOPAS software \cite{Coelho2018}, while thermal expansion diffraction data were obtained using GSAS-II software \cite{toby2001expgui}.

\paragraph{Simulation methods USPEX and Quantum Espresso}
Structural predictions were carried out using the USPEX (Universal Structure Predictor: Evolutionary Xtallography) algorithm, a powerful evolutionary algorithm capable of identifying stable crystal structures for a given stoichiometry \cite{oganov2006, lyakhov2013, oganov2011}. In this study, the structural exploration on the \ce{H2O}-\ce{N2} system, treats \ce{H2O} and \ce{N2} molecules as unbreakable building blocks from which candidate structures were generated.\\
USPEX initially produces randomised structures under chemical and physical constraints, including fixed stoichiometry, predefined inter- and intra-molecular distances, to ensure chemically reasonable starting configurations. These candidate structures were then subjected to a multistage relaxation protocol using Quantum ESPRESSO \cite{Giannozzi2009, Giannozzi2017}, involving four consecutive relaxation steps with systematically increasing computational accuracy, until forces were lower than 0.5 mRy/atom, and stresses were within 1.0 kbar of the target pressure. \\
During relaxation, any structure that became chemically unstable, such as through molecular dissociation or violation of the imposed distance constraints, was discarded. At the end of each generation, the structures with the lowest free energy were selected as the basis for the next generation. USPEX employs various evolutionary operators (e.g., heredity, mutation, and substitution) to generate new structures based on these best-performing candidates.

These structural searches were performed on fixed compositions, allowing for direct comparison of structural stability. Final candidate structures were evaluated using the convex hull construction to determine their thermodynamic stability relative to competing phases.

The local structural relaxations and computation of total enthalpies were performed using Quantum ESPRESSO \cite{Giannozzi2009, Giannozzi2017}, using progressively tighter constraints on the forces and stresses, and better convergence parameters. The Kohn-Sham wavefunctions were expanded on a plane-wave basis of up to 80 Ry, and integration over the Brillouin zone was performed using a k-space point density of 1.0053 up to 0.8796 \AA$^{-1}$. We employed the Optimized Norm-Conserving Vanderbilt pseudopotentials (ONCV) from pseudo-dojo. The exchange-correlation functional was treated within the Perdew-Burke-Ernzerhof (PBE) approximation. 

Chosen crystal structures were then further relaxed as a function of pressure from 0 to 100 GPa. For these calculations we employed a cutoff of 90 Ry, k-point density of 0.185 \AA$^{-1}$, and a fixed occupation.

\bibliographystyle{ieeetr}
\bibliography{Biblio}

\section{Acknowledgments}
We acknowledge the Institut Laue-Langevin (ILL) for providing beamtime on D20 (proposal number 5-22-834). The authors acknowledge assistance from Claude Payre and James Maurice with the high pressure set up. S.E. and S.D.C. acknowledges computational resources from ...... S.E. and L.E.B. acknowledges the financial support by the European Union - NextGenerationEU (PRIN N. 2022NRBLPT). G.T. and L.E.B. acknowledges the financial support by the ANR-23-CE30-0034 EXOTIC-ICE. T.P., R. G. and L.E.B. acknowledge the financial support from the Swiss National Fund (FNS) under Grant No. 212889. S.B. thanks the CNR-FOE-LENS.

\section{Data Availability Statement}

The experimental data supporting this work  phases are available at the ILL data repository under the following DOI: https://doi.ill.fr/10.5291/ILL-DATA.5-22-834 .
These datasets are currently under embargo until 2027. However, they are available from the authors upon reasonable request.

\end{document}